\documentclass[preprint,12pt]{elsarticle}
\usepackage{algorithm}
\usepackage{algpseudocode}
\usepackage{amsmath}
\usepackage{subfigure}
\usepackage[colorlinks]{hyperref}

\usepackage{amssymb}

\journal{}

\begin{document}

\begin{frontmatter}



\title{Development of Maximum Conical Shock Angle Limit for Osculating Cone Waveriders}


\author[inst1]{Agnivo Ghosh}

\affiliation[inst1]{organization={Department of Aerospace Engineering},
            addressline={Indian Isntitute of Science}, 
            city={Bengaluru},
            postcode={560012}, 
            state={Karnataka},
            country={India}}

\author[inst1]{Srisha M V Rao*}
\begin{abstract}
Hypersonic waveriders are special shapes with leading edges coincident with the body’s shock wave, yielding high lift-to-drag ratios. The waverider geometry results from streamline tracing using the solutions of a basic flow field such as the wedge or the cone for specified shock and base curves. The base and shock curves can be independently prescribed in the osculating cone method enabling a larger design space. Generally, low values of the conical shock angle ($9^\circ-15^\circ$) are used. The lack of any method to limit the maximum cone angle for osculating cone waverider motivates this study. Mathematical expressions are derived for geometrical conditions that result in successful osculating cone waverider generation. A power law curve and a B\'{e}zier curve are analyzed. Closed-form expressions for the maximum cone shock angle are obtained for the power law curve. A numerical procedure to solve the same for the B\'{e}zier curve is developed. The results, for a typical Mach number of 6.0, evidently show that the maximum cone shock angle for successful waverider generation is significantly lower than the maximum angle for attached shock solutions. The limiting conditions developed will be essential in constraining the waverider sample space for automated multi-objective optimization routines.
\end{abstract}

\begin{graphicalabstract}
\includegraphics[width=\textwidth]{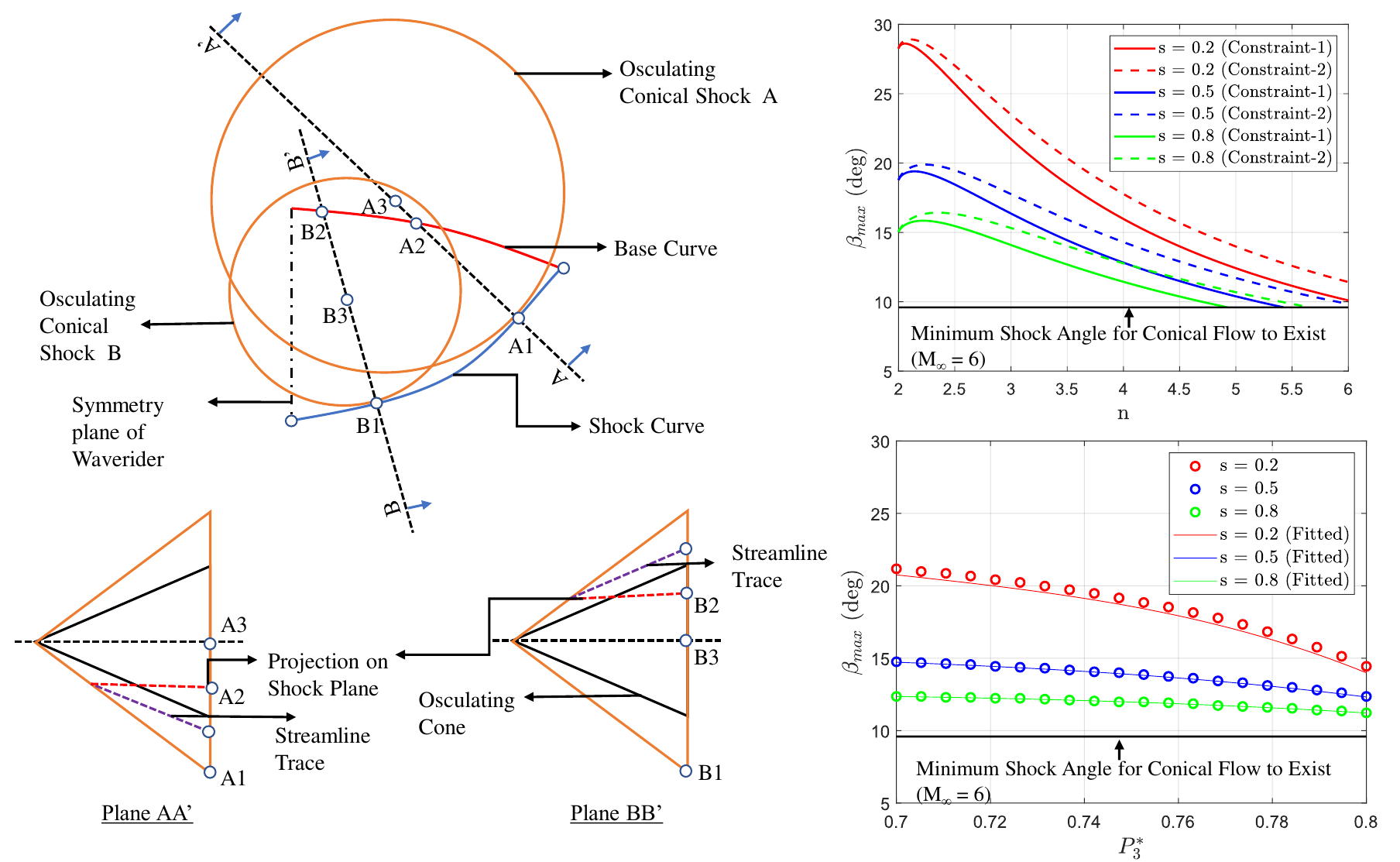}
\end{graphicalabstract}

\begin{highlights}
\item In literature, researchers have used low values of conical shock angle without any clear justification. 
\item The maximum conical shock angle limit is derived for a generic shock curve.
\item Two specific shock curves are prescribed and limits are established by analytical and numerical means.
\item Trends show that the maximum conical shock angle is highly sensitive to the design parameters of the shock curve.
\end{highlights}


\end{frontmatter}


\section{Introduction}
\label{Intro}
Sustained long range atmospheric flight vehicle in the hypersonic regime is under active development for military and access to space applications \cite{Anderson2019HypersonicEdition,Bertin1994HypersonicAerothermodynamics}. Air-breathing propulsion system (scramjet) is preferred considering the need to carry higher payload. Unique challenges of shock waves, aerothermal loading and advantages of ram- compression demand an integrated vehicle design philosophy. Typical terrestrial re-entry vehicles \cite{Berry2009AerothermalVehicle,Vago2015ESAMars}utilised blunt aerodynamic shapes which reduced thermal loading and provided sufficient drag for aerobraking. However, for long range hypersonic cruise vehicles lift is equally significant in addition to mitigation of aero-thermal loads and high drag. Conventional blunt hypersonic vehicle configurations showed an upper limit of lift to drag ratio, which prompted Kuchemann \cite{Kuchemann1965HypersonicProblems} to propose the concept of “lift to drag ratio barrier “. Researchers sought for special aerodynamic shapes capable of breaking the lift to drag ratio barrier . The concept of waverider was first introduced by Nonweiler \cite{Nonweiler1959AerodynamicVehicles} in 1959. A standard waverider consists of an upper freestream surface and a lower compression surface. They are designed in such a way that the shock remains completely attached to the leading edge. In essence, it seems as though the vehicle is riding on the shock.  In doing so, it can encapsulate a higher-pressure region under the lower surface thereby generating higher lift, hence breaking the aerodynamic barrier.

In practice, waveriders are generated by an inverse design methodology which requires an assumption of a basic flowfield. The upper surface of a waverider is developed by projecting the base curve to the shock plane and subsequently carrying out streamline tracing \cite{Billig2000StreamlineVehicles}.The intersection of the upper freestream surface and the compression surface forms the leading edge of a waverider. The earliest design by Nonweiler \cite{Nonweiler1959AerodynamicVehicles} used flow past a wedge as the basic flowfield. It is known as the caret ($\Lambda$) based waveriders. The caret waveriders were designed considering the base curve to be straight lines. This was further modified by Starkey et al \cite{Starkey1998AWaveriders} who used power law as the base curve. The development of these type of waveriders were computationally inexpensive as they involved solution of oblique shock equations which are algebraic in nature. Despite breaking the ‘L/D barrier’ these waveriders were too thin to have sufficient volume which is a critical consideration in terms of the payload of a vehicle. Thus, there was a need to develop waveriders which can provide sufficient volumetric efficiencies($\eta$) in addition to aerodynamic efficiency.A separate class of waveriders emerged with conical flow as the basic flow field. These required the solution of the Taylor Maccoll equations as opposed to the oblique shock equations that were utilised to develop caret waveriders. Jones et al \cite{Jones1968AFields} in 1968, generated a simple cone derived waverider from axisymmetric flow fields. Subsequently, Rasmussen et al \cite{Rasmussen1980WaveriderCones} developed waveriders with inclined circular and elliptical cones as generating bodies using hypersonic small perturbation theory, but they were only limited to small cone angles. Goonko et al \cite{Goonko2000Convergent-Flow-DerivedWaveriders} developed a special type of waveriders utilising axisymmetric flows inside constricting ducts specifically conical trumpet ducts. Cone-based waveriders offered better geometric features, but once a Mach number and shock angle were chosen, only the upper surface on the base plane could be adjusted.

Increasing competitive objectives necessitated a shift to look at waveriders capable of providing an even larger design space. One such development was first proposed by Sobiciezky et al \cite{Sobieczky1990HypersonicWaves.} in 1990 as the osculating cone theory. It was further improvised by Rodi \cite{Rodi2005TheGeneration} and became famously known as the osculating flow field method. It eased the requirement for the flowfield at each osculating plane to be conical. The osculating cone approach gives the freedom to select a specific shock curve at the base plane, with the only requirement that it must be second order continuous. . Despite this being not an actual flow field, the approximation of local conical flow field sufficiently improved the design space which has become a precursor for optimisation against several competing objectives. Accurate shape optimisation requires high fidelity CFD analysis, which in turn necessitates rapid geometry generation of waveriders. Streamline tracing in cone derived waveriders is a computationally expensive procedure primarily due to the requirement of the solution of a system of differential equations.Ding et al \cite{Ding2015SimplifiedWaverider} proposed the simplified osculating cone waverider, which drastically reduced the computational cost of developing a waverider. Even though these waveriders had better volumetric efficiencies, spillage was observed at the tip of the waverider which decreased the aerodynamic efficiency. On similar grounds, recently a novel Artificial Neural Network based approach to simplify streamline tracing was put forward by the authors' group \cite{Rao2023ArtificialDesign}. Modifications have also been done on osculating cone theory for a variety of unique competing objectives.For example, Chen et al \cite{Chen2019AWaveriders} developed a unique design methodology in line with the osculating cone theory to improve the volumetric aspect of waveriders. Recently, researchers have also looked at bulging the upper surface of a waverider to increase the volume of a waverider at the cost of aerodynamic performance \cite{Liu2020EffectPerformances}.Another limitation of the waverider design is that they are designed for a particular cruise Mach number. However, in practical situations there is a requirement of forebodies to perform well over a wide range of speeds. To address this issue many modifications have been done to the shock curve \cite{Liu2019NovelNumber} or by changing the shape of the leading edge \cite{Li2018DesignRange} to name a few. \cite{Ding2017AnMethodology,Zhao2020AnConfiguration} summarises a review of the most recent cutting-edge waverider design methods.

In addition to their aerodynamic performance, Osculating Cone Waverider as forebodies have showed promising results required for inlet integration \cite{OBrien2001Rocket-BasedVehicle,He2016DesignInlet}. With osculating cone waveriders it is possible to choose a specific area of uniform flow which is required for inlet integration. Despite all these advantages, development of waveriders for practical usage is a difficult proposition owing to challenges in manufacturing and high heat transfer rates at the sharp leading edge.Therefore, to mitigate the problem of high heat transfer rates at the leading edge, researchers have looked at blunting the leading edge \cite{Santos2009BluntnessFlow,Gillum1997ExperimentalEdges}of the waveriders, using counter-flowing jets \cite{Li2016ResearchRadii} to name a few. Chen et al \cite{Chen2011BluntnessWaverider} studied the bluntness impact on a cone derived waverider numerically and concluded that even though blunting the leading edge partially alleviates the problems associated to thermal protection, however the aerodynamic performance suffers greatly. On similar lines, experiments as well as numerical simulations were carried out by Liu et al \cite{Liu2017EffectAttack} on a blunt waverider under different angle of attack. Novel blunting methods \cite{Li2017AerodynamicFlows,Qu2021NumericalMethod} have also been aimed to improve aerodynamic performance alongside reducing aerothermal loads.

Nowadays, optimisation studies for vehicle design is a standard procedure for selection of test model \cite{Singh2017OptimizationTLBO,Parsonage2023AOptimisation,Eyi2018AerothermodynamicVehicles}. For accurate optimisation, there is a requirement of large dataset which in turn requires accurate range of design variables. One of the foremost parameters which influence the shape of a waverider is the shock angle ($\beta$). Typically, higher values of $\beta$ will lead to waveriders with higher volumetric efficiencies but with low aerodynamic efficiencies. Thus, for multi-objective optimisations it is necessary to include the entire range possible for each design variable. Moreover, optimisation algorithms like NSGA-II (Non-dominated Sorting Algorithm) \cite{Deb2002ANSGA-II}, TLBO (Teaching learning-based Optimization)\cite{Rao2016TeachingAlgorithm} require an automated framework which involves geometry generation followed by CFD analysis, hence proper understanding of design limits are required in order to avoid development of infeasible geometries. Theoretically, as per the standard osculating cone theory, waveriders can be developed with shock angles varying between the minimum angle required ($\sin^{-1}(1/M_{\infty}$)) for conical flow to exist to the shock detachment angle. A comprehensive study of the conical shock angle used in open literature alongside the respective shock detachment angle is shown in Table \ref{Review_Table}. From Table \ref{Review_Table}, it is evident that even though $\beta$ can be chosen to be as high the shock detachment angle, still authors have tended to choose very low values without clear justification of the same.In addition to this, an optimisation study \cite{Chen2008Multi-ObjectCone} was performed on osculating cone waverider, however no proper range of $\beta$ was provided. 

The problems that restrict the development of osculating cone waveriders have been very briefly stated in Kontogiannis et al \cite{Kontogiannis2017EfficientGeometries} and Son et al \cite{Son2022AMethods}, but an explicit range for the shock angle was not apparent.However, no analytical expression/ numerical data is available in open literature which limits the formation of osculating cone waveriders. One of the primary requirements of an osculating cone waverider is that the shock curve at the base plane should be second order continuous, making power law and B\'{e}zier curves to be ideal solutions. With this motivation in mind the article focuses on limiting the shock angle of osculating cone waveriders derived from power law and B\'{e}zier curves.
\begin{table}[hbt!]
\caption{\label{Review_Table} Review of Shock Angles Used in Open Literature}
\centering
\begin{tabular}{|c|c|c|c|}
\hline
Ref & Mach Number & $\beta$ Used & $\beta_d$\\
\hline
Qing et al \cite{Xiao-qing2009OrthogonalityWaverider}  & 10 & $9^\circ-15^\circ$ & $73.41^\circ$\\
\hline
Chen et al \cite{Chen2019AWaveriders} & 6 & $12^\circ$ & $72.54^\circ$\\
\hline
Chen et al \cite{Chen2020DesignAngles} & 6 & $12^\circ$ & $72.54^\circ$\\
\hline
Chen et al \cite{Chen2019WaveriderAngle} & 6 & $10^\circ-14^\circ$ & $72.54^\circ$\\
\hline
\end{tabular}
\end{table}
\section{Standard Osculating Cone Waverider Design Methodology}
\begin{figure}[H]
\subfigure[Base View]{\includegraphics[width=\textwidth]{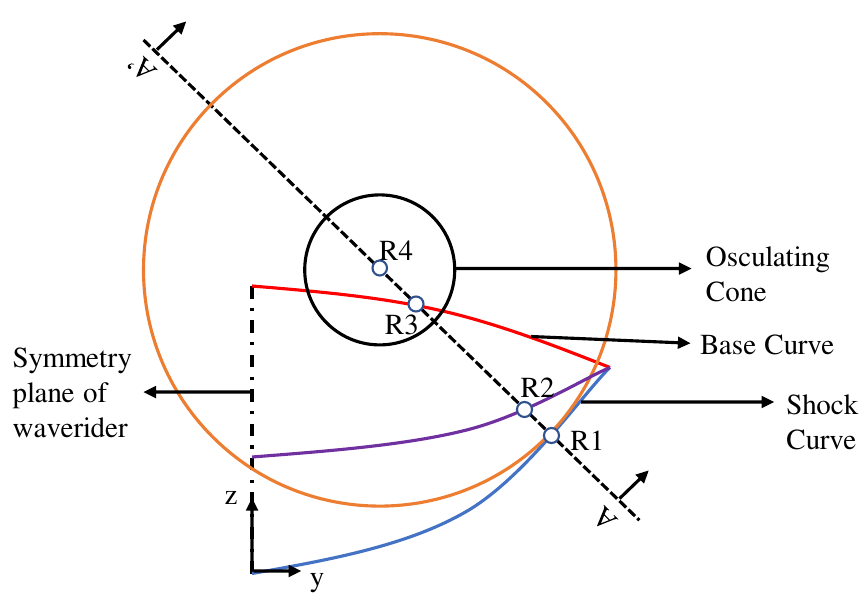}}
\subfigure[Plane AA']{\includegraphics[width=\textwidth]{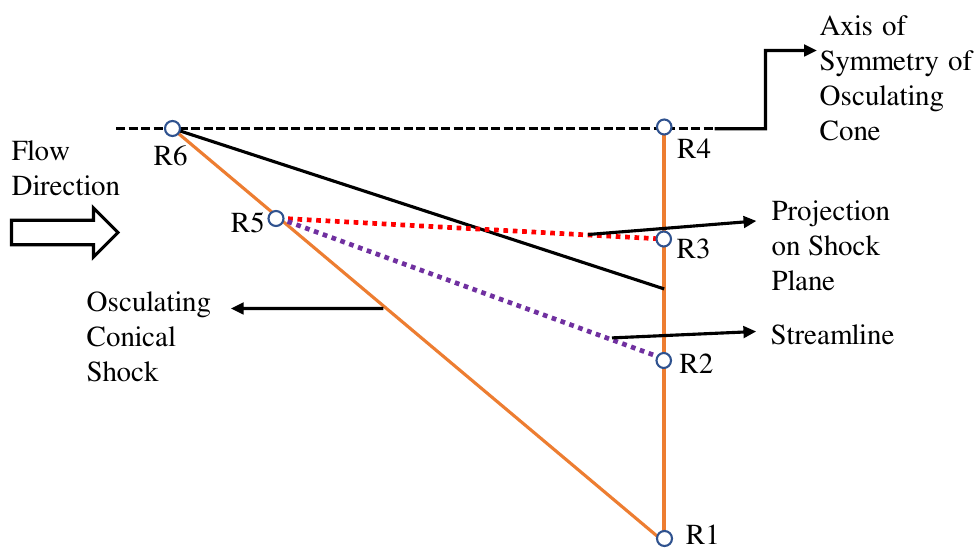}}
\caption{Design Methodology of Osculating Cone Waverider}
\label{Design_Method}
\end{figure}

The general standard osculating cone waverider design methodology with predefined shock and base curve at the base plane has been adopted in the current study. The steps followed in the development of the osculating cone waverider has been shown in Algorithm [\ref{alg:cap}]. In Figure [\ref{Design_Method}], the base curve, shock curve and the compression surface curve at the base plane is shown by red , blue and violet colours respectively . $R1$ represents any general point on the predefined shock curve at the base plane. A local osculating plane is created at $R1$ centred at $R4$ which intersects the base curve at $R3$. For the development of the compression surface streamline tracing is done by numerically solving Taylor Maccoll equations. $R2$ corresponds to the compression surface point at the base plane.
\label{Standard_Design}
\begin{algorithm}
\caption{Design Procedure for Osculating Cone Waverider}\label{alg:cap}
\begin{algorithmic}
\State Inputs- Mach Number, Shock Angle ($\beta$) or Cone Angle ($\theta$), Base Curve $z_b=g(y)$ and Shock Curve $z_s=f(y)$
\If{$M$ and $\theta$ is given}
    \State Calculate $\beta$ using Conical Shock Theory
\Else
    \State Use $M$ and $\beta$
\EndIf
\State Step 1: Solve Taylor Maccoll Equation to find $V_r$ and $V_\theta$.
\State Step 2: Discretise the Shock Curve and Develop Osculating Planes.
\State Step 3: Find the intersection of the osculating planes with the base curve. 
\State Step 4: Project each Point found in Step 3 to the Shock Plane to form the Upper Surface of Waverider.
\State Step 5: Join the points on the shock plane to form the Leading Edge of the Waverider.
\State Step 6: Carry out Streamline Tracing from the points on the shock plane to form the Lower Compression Surface. 
\end{algorithmic}
\end{algorithm}
\section{Constraint Development for a Waverider with a Generic Shock Curve }
\label{Section_Gen_Constraint}
The successful geometry generation (as discussed in Sections [\ref{Intro} , \ref{Standard_Design}]) of an osculating cone waverider demands a careful selection of the shock and base curves at the base plane. Besides minimum requirements on continuity of the curve, it is found that the following two conditions should be satisfied while selecting a shock curve.
\begin{enumerate}
    \item \textbf{\underline{Condition 1}}- Radius of curvature ($R$) of the shock curve at the base plane is more than the distance between the shock curve point and the base curve measured along the osculating plane ($H$).
    \item \textbf{\underline{Condition 2}}- The osculating planes do not intersect between the base curve and the shock curve.
\end{enumerate}
\begin{figure}[H]
\subfigure[Base View]{\includegraphics[width=\textwidth]{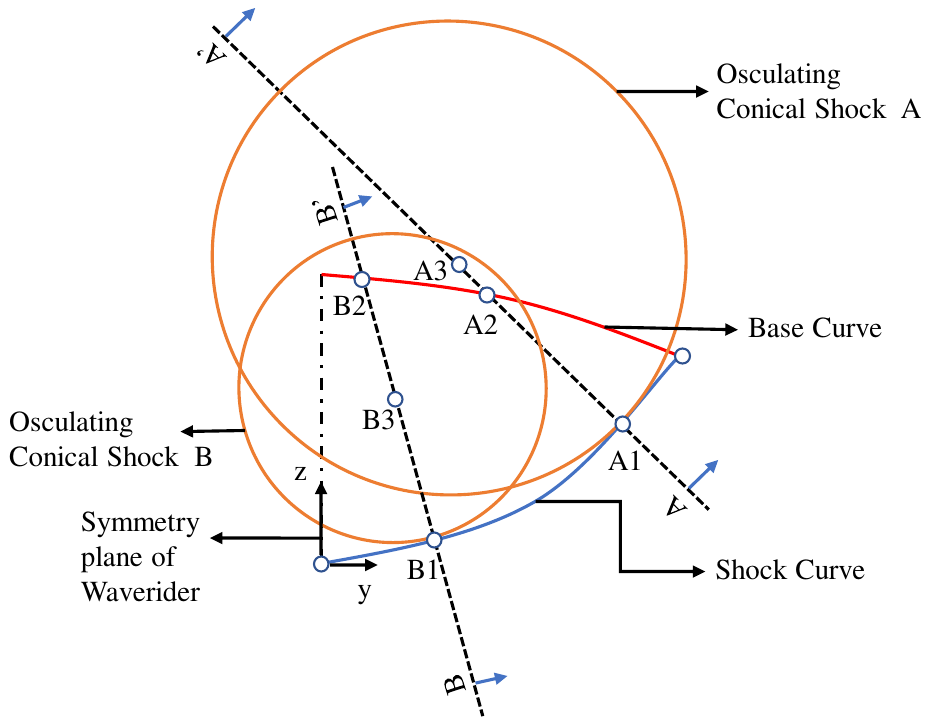}\label{fig13a}}
\subfigure[Plane View]{\includegraphics[width=\textwidth]{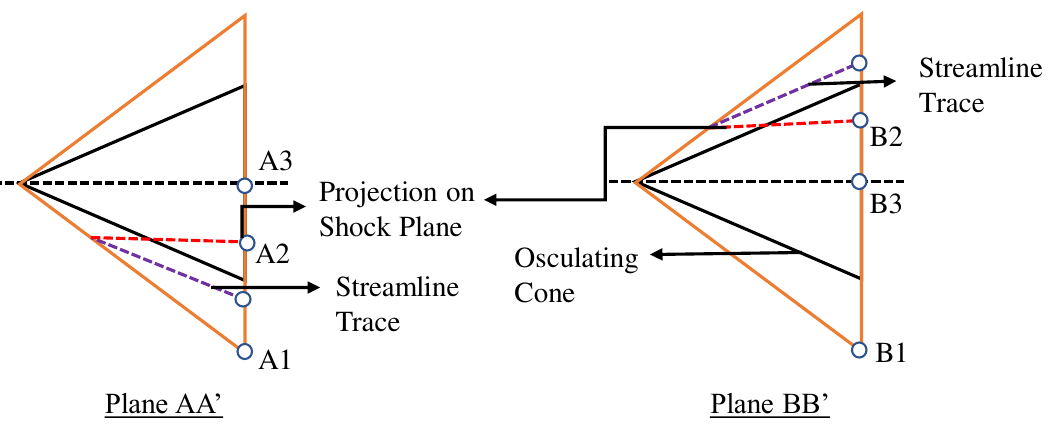}\label{fig13b}}
\caption{Pictorial Description of Constraint-1}
\label{Constraint_1}
\end{figure}
Condition 1 is explained with the help of illustrations in  Figure [\ref{Constraint_1}]. Two osculating planes (AA' and BB') are shown, among which the plane  AA' satisfies Condition 1 whereas BB' violates Condition 1. On the plane AA' , $R=(A3-A1)$, $H=(A2-A1)$, and it is evident that $R>H$. Thus, the lower compression surface along AA' is successfully generated by streamline tracing as is shown in the Plane AA' view of Figure [\ref{Constraint_1}]. However, in the case of Plane BB' ,$R=(B3-B1)$, $H=(B2-B1)$, and $R < H$. Here, streamline tracing can only be possible in a direction which forms the trace above the freestream surface which doesn't yield the appropriate compression surface. This has been shown in the Plane BB' view of Figure [\ref{Constraint_1}].

\begin{figure}[hbt!]
    \centering
    \includegraphics[width=0.8\textwidth]{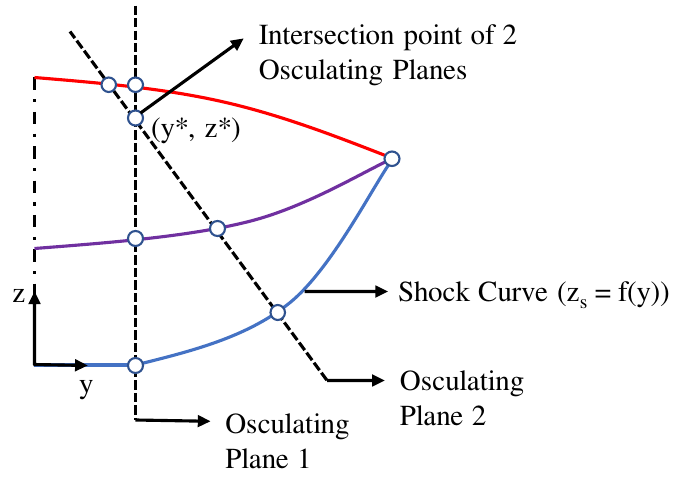}
    \caption{Pictorial Description of Constraint-2}
    \label{Constraint_2}
\end{figure}

Figure \ref{Constraint_2} shows the case where, two osculating planes (Osculating Planes 1 and 2) intersect between the base and the shock curve (at the point $(y^* , z^*)$) which is a violation of Condition 2. This would lead to disrupted upper surface at the base plane as well as intersecting streamlines which is not a feasible proposition.

If it is ensured that the minimum radius of curvature ($R_{min}$) of the shock curve is greater than the maximum gap between the base and the shock curve ($H_{max}$), then conservatively, it is expected that all points of the shock curve satisfy Condition 1. Moreover, from the nature of the shock curves, it is quite visible that ensuring consecutive osculating planes don’t intersect within the waverider is a good enough criterion for assuring that none of the osculating planes intersect between the shock curve and the base curve. Hence, to summarise,

\textbf{\underline{Constraint 1}}- $R_{min} > H_{max}$

\textbf{\underline{Constraint 2}}- $z*_{min} > H_{max}$

The exact mathematical condition for Constraint 1 and Constraint 2 on an arbitrary shock curve $z=f(y)$ is derived in Appendix. It is evident that solving $3f''^2f'-f'^2f'''-f'''= 0$ will lead to the particular value of $y$ which is most susceptible to fail for both the constraints [Equations \ref{Final Equation 1},\ref{Final Equation 2}]. Substituting the corresponding value of $y$ in the expression of $‘R’$ [See Equation\ref{Radius_Curvature_equation}] and $‘z*’$ [See Equation \ref{z*_Equation}] will result in $R_{min}$ and $z^*_{min}$ respectively, both of which should be greater than $H_{max}=L \tan\beta$ for the development of a feasible waverider.

\section{Selection of Shock Curve}
In this section, two different shock curves are prescribed that are atleast $C^2$ continuous (Second derivatives must be continuous) which are suitable candidates for $z_s=f(y)$ described in Section [\ref{Section_Gen_Constraint}]. $C^2$ continuity is an essential requirement since the radius of curvature at a location on the shock curve needs to be calculated. Therefore, smooth curves such as Power law and B\'{e}zier curves are excellent choices.
\subsection{Power Law}
\begin{figure}[hbt!]
    \centering
    \includegraphics[width=\textwidth]{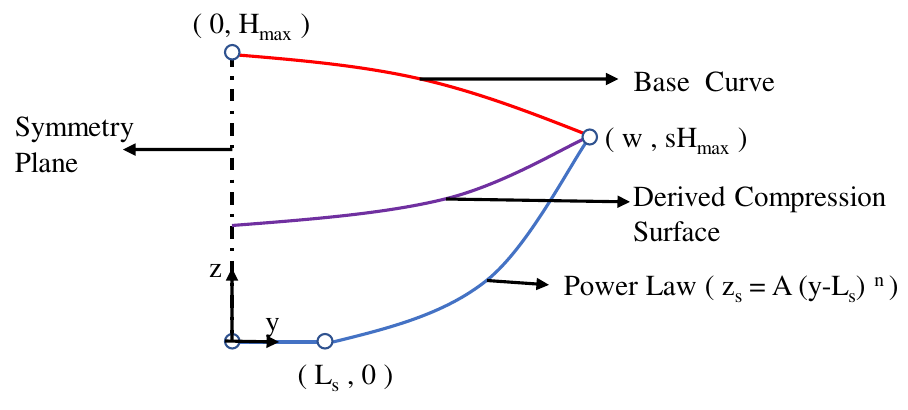}
    \caption{Power Law shock curve description at the base plane}
    \label{Power_Law_Basic}
\end{figure}

Figure \ref{Power_Law_Basic} shows the shape of the compression curve at the base plane derived from a power law shock curve and an arbitrary base curve ($z_b=g(y)$) specified at the base plane. The general equation of the power law shock curve is given by-
\begin{equation}
\begin{split}
    \label{Powerlaw_GeneralEquation}
    &z_s=0 \quad \textrm{for} \quad 0 \leq y \leq L_s\\
   &z_s = A(y-L_s)^n  \quad \textrm{for} \quad L_s \leq y \leq w
    \end{split}
\end{equation}
Where,
\begin{equation}
    A = \frac{sH_{max}}{(w-L_s)^n}
\end{equation}
The parameters on which the shock curve depends are-
\begin{enumerate}
    \item  s - Ratio of tip height of waverider at base plane to $H_{max}$
    \item $L_s$ - Straight Portion of the Shock Curve 
    \item $\beta$ - Conical Shock Angle 
    \item $n$ - Exponent of the Power Law
    \item $w$ - Semi Span of the Waverider
\end{enumerate}
\subsection{B\'{e}zier Curve}
\begin{figure}[hbt!]
    \centering
    \includegraphics[width=\textwidth]{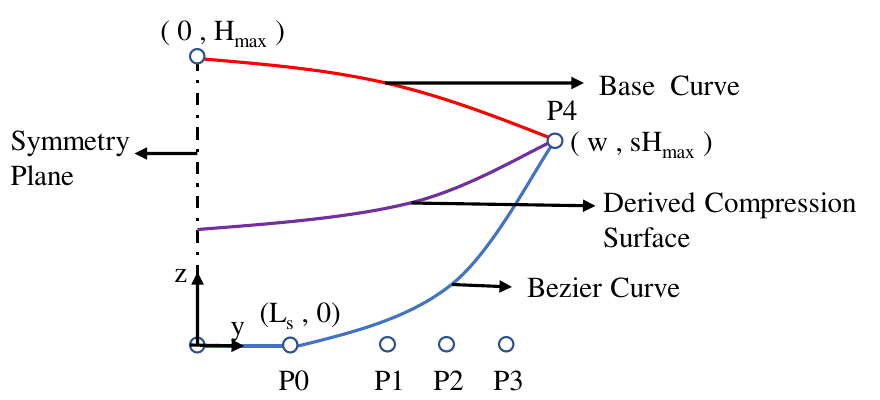}
    \caption{B\'{e}zier shock curve description at the base plane}
    \label{Bezier_Basic}
\end{figure}

Figure \ref{Bezier_Basic} shows the shape of the compression curve at the base plane derived from a B\'{e}zier shock curve with three control points and an arbitrary base curve ($z_b=g(y)$) specified at the base plane. B\'{e}zier curves are parametric curves which are defined using Bernstein basis polynomials. The generalised expression of a B\'{e}zier curve is given by- 
\begin{equation}
    B(t) = \sum_{i=0}^{n} b_{i,n}(t)P_i \quad \textrm{for} \quad  0\leq t \leq 1
\end{equation}
Where,
\begin{equation}
    b_{i,n}(t)= \binom{n}{i}t^i(1-t)^{n-i} \quad \textrm{for} \quad i=0,1....n
\end{equation}
And,
\begin{center}
 $P_i$ are the control points.
 \end{center}
 The straight portion of the shock curve is given by- 
 \begin{equation}
         z_s = 0 \quad \textrm{for} \quad 0 \leq y \leq L_s
 \end{equation}
 The expression for the B\'{e}zier part of the  shock curve is given by-
  \begin{multline}
    y_s = (1-t)^4P_{0y} + 4t(1-t)^3P_{1y} +6t^2(1-t)^2P_{2y} +4t^3(1-t)P_{3y}+t^4P_{4y} \\\quad \textrm{for} \quad 0 \leq t \leq 1
\end{multline}
\begin{multline}
    z_s = (1-t)^4P_{0z} + 4t(1-t)^3P_{1z} +6t^2(1-t)^2P_{2z} +4t^3(1-t)P_{3z}+t^4P_{4z} \\\quad \textrm{for} \quad 0 \leq t \leq 1
\end{multline}
 Where, $P_{0y},P_{1y}...P_{4y}$ and $P_{0z},P_{1z}...P_{4z}$ represent the $y$ and $z$ coordinates of the control points $P_0,P_1...P_4$ respectively. The coordinates of end control points $P_0$ and $P_4$ are given by $(0,L_s)$ and $(w,sH_{max})$ respectively.
\\
\begin{figure}[hbt!]
\centering
\subfigure[Base View]{\includegraphics[width=0.8\textwidth]{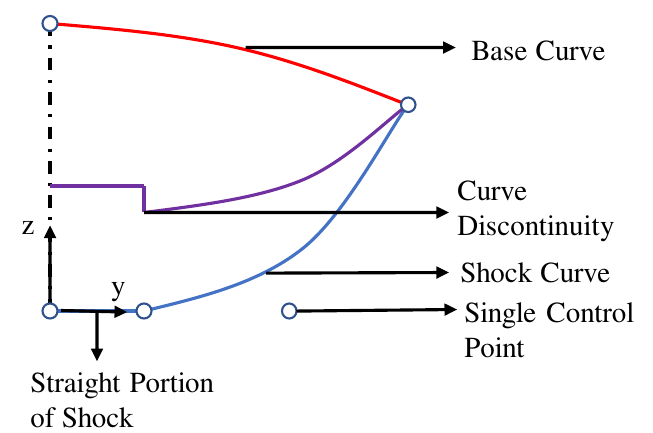}}
\subfigure[Plane View]{\includegraphics[width=0.8\textwidth]{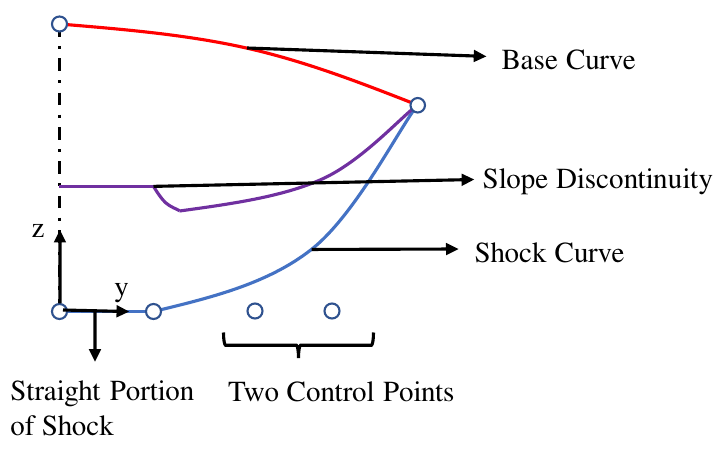}}
\caption{Waveriders derived from a b\'{e}zier shock curve with one and two intermediate control points}
\label{Effect_of_3}
\end{figure}
 Figure \ref{Effect_of_3} shows the typical shape of the waverider at the base plane with one and two intermediate control points. It is evident that with a single control point, there is an existence of point discontinuity on the compression curve, whereas with two intermediate control points slope discontinuity exists. Thus, in order to develop a waverider with smooth compression surface, a B\'{e}zier curve of minimum three intermediate control points is necessary. The same is used for subsequent studies in this article.
 \section{Application of Constraint on Selected Shock Curves}
 \subsection{Power Law}
 Application of Constraint 1 and Constraint 2 in the case of power law shock curve led to inequalities \ref{Constraint_1_Equation} and \ref{Constraint_2_Equation} respectively.

 \begin{equation}
    \label{Constraint_1_Equation}
    \frac{3^{1.5} (n-1)^{0.5} (w-L_s)^{\frac{n}{n-1}}}{(2n-1)^{\frac{2n-1}{2n-2}} s^{\frac{1}{n-1}} n^{\frac{1}{n-1}} (n-2)^{\frac{n-2}{2n-2}}} > (tan\beta)^{\frac{n}{n-1}}
\end{equation}
\begin{multline}
    \label{Constraint_2_Equation}
    \frac{(w-L_s)^{\frac{n}{n-1}}}{s^{\frac{1}{n-1}}} \Biggl[\Bigl(\frac{3}{(2n-1)^{\frac{n}{2n-2}}(n-2)^{\frac{n-2}{2n-2}}n^{\frac{1}{n-1}}}\Bigr) + \Bigl( \frac{(n-2)^{\frac{n}{2n-2}}}{n^{\frac{n}{n-1}}(2n-1)^{\frac{n-2}{2n-2}}}\Bigr) \Biggr] \\> \Bigl(tan\beta\Bigr)^{\frac{n}{n-1}}
\end{multline}
It can be seen clearly from inequalities \ref{Constraint_1_Equation} that $s \neq 0$ and $n > 2$ which further justifies that the shock curve should have atleast $C^2$ continuity. Moreover, both the inequalities \ref{Constraint_1_Equation} and \ref{Constraint_2_Equation} yield an upper limit to the conical shock angle ($\beta$) for a given set of design parameters.Results show that the maximum shock angle $(\beta_{max})$ is significantly less than the shock detachment angle $(\beta_d)$ for combinations of different design parameters.The minimum and maximum shock angles are computed for a typical hypersonic Mach number of 6. A parametric study was carried out to understand the effect of different design variables on the upper limit of conical shock angle which is elucidated in the following 3 cases.
\begin{enumerate}
    \item \underline{Case-1} - The effect of exponent of the power law was studied for 3 distinct cases of tip height($s$). Three values of $s$ lying between 0.2 to 0.8 are chosen. It can be seen from Figure[\ref{Power_Graph_1}] that the maximum shock angle for successful waverider generation $(\beta_{max})$ increases with decrease in $s$. This sort of behaviour can be attributed to the fact that  the radius of curvature increases with decrease in $s$. Moreover, it is evident from Figure [\ref{Power_Graph_1}] that for any value of $s$, the maximum permissible shock angle($\beta_{max}$) reaches a maximum around $n=2.1$, after which it decreases steadily.
    \begin{figure}[hbt!]
    \centering
    \includegraphics[width=0.9\textwidth]{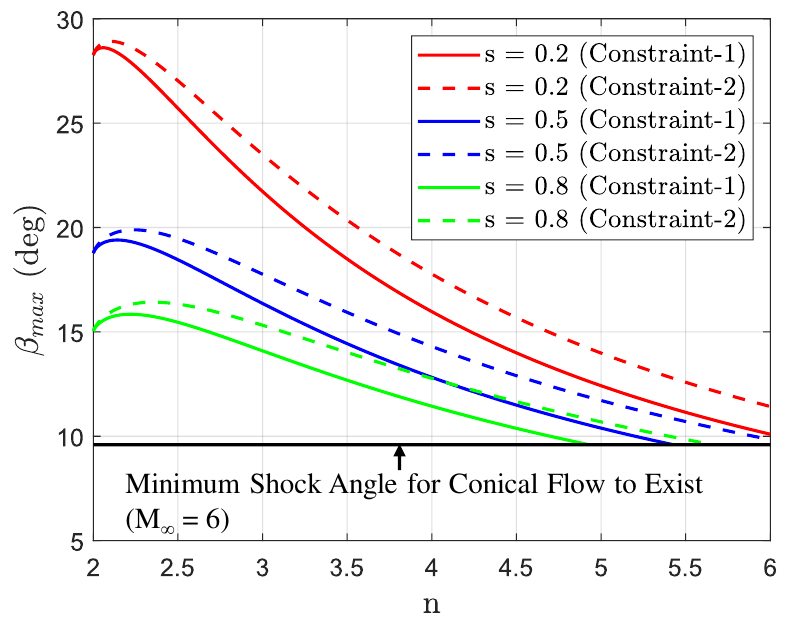}
    \caption{Maximum Permissible shock angle$(\beta_{max})$ vs n for 3 different s}
    \label{Power_Graph_1}
    \end{figure}
    \item \underline{Case-2} - Figure[\ref{Power_Graph_2}] plots the effect of exponent of the power law for 3 distinct cases of straight portion of the shock curve($L_s^* = L_s/w$) which correspond to 10\%,15\% and 20\% of the semi span of the waverider respectively .Similar sort of trend is observed as in the case of variation with $s$.
    \begin{figure}[hbt!]
    \centering
    \includegraphics[width=0.9\textwidth]{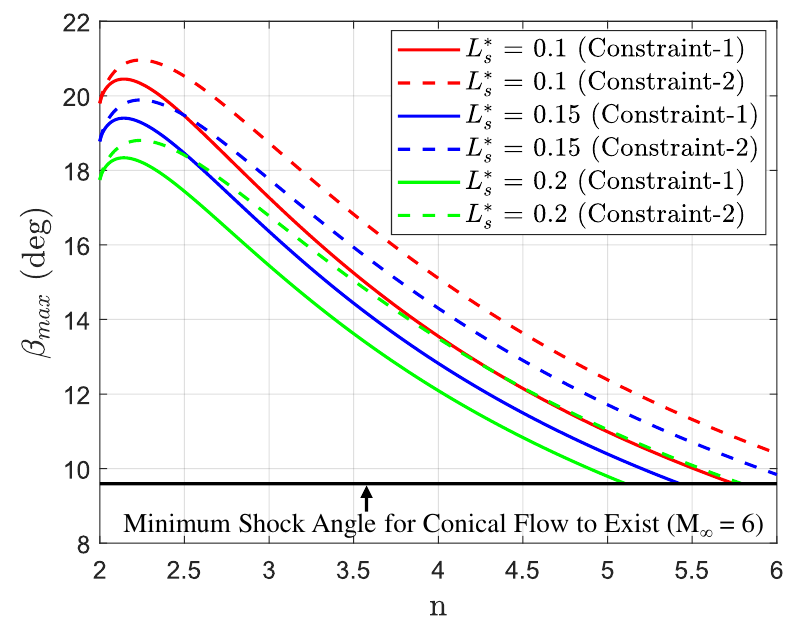}
    \caption{Maximum Permissible shock angle$(\beta_{max})$ vs n for 3 different $L_s^*$}
    \label{Power_Graph_2}
    \end{figure}
    \item \underline{Case-3} - The effect of tip height of the waverider was studied for 4 distinct cases of the exponent of the power law ($n$). It can be seen from Figure[\ref{Power_Graph_3}] that when $s\approx0$ the maximum permissible shock angle approaches conical shock detachment angle $(\beta_d)$. The rate of drop of  $\beta_{max}$ with $s$ increases with increase in $n$. For example, the rate of drop is so high in case of $n=6$ that no osculating cone waverider can be formed for $s>0.3$ even for very low shock angles. Hence, in order to utilise the full design space one should typically avoid higher values of $n$.
    \begin{figure}[hbt!]
    \centering
    \includegraphics[width=0.9\textwidth]{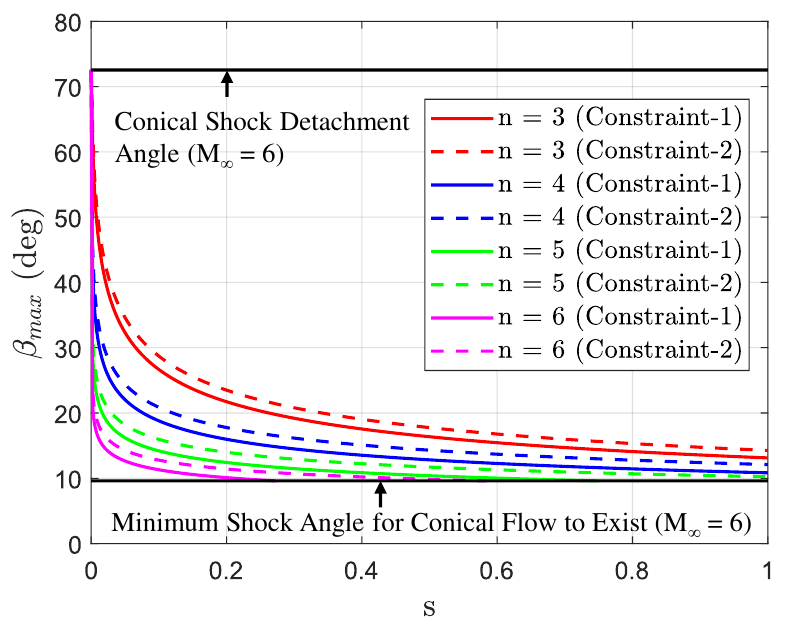}
    \caption{Maximum Permissible shock angle$(\beta_{max})$ vs s for 5 different n}
    \label{Power_Graph_3}
    \end{figure}
\end{enumerate}

Another important observation which can be made from equations[\ref{Constraint_1_Equation},\ref{Constraint_2_Equation}] that, larger the width of the vehicle higher is the minimum radius of curvature of the shock curve. This will in turn increase the maximum shock angle $(\beta_{max})$ for a feasible design. This can be a crucial consideration while selecting the aspect ratio $(w/L)$ of a waverider. All the graphs [\ref{Power_Graph_1},\ref{Power_Graph_2},\ref{Power_Graph_3}] shown in this article are corresponding to aspect ratio ($w/L = 0.4$). Similar sort of trends are observed with other aspect ratios.

\subsection{B\'{e}zier Curve}
\begin{figure}[hbt!]
    \centering
    \includegraphics[scale=0.8]{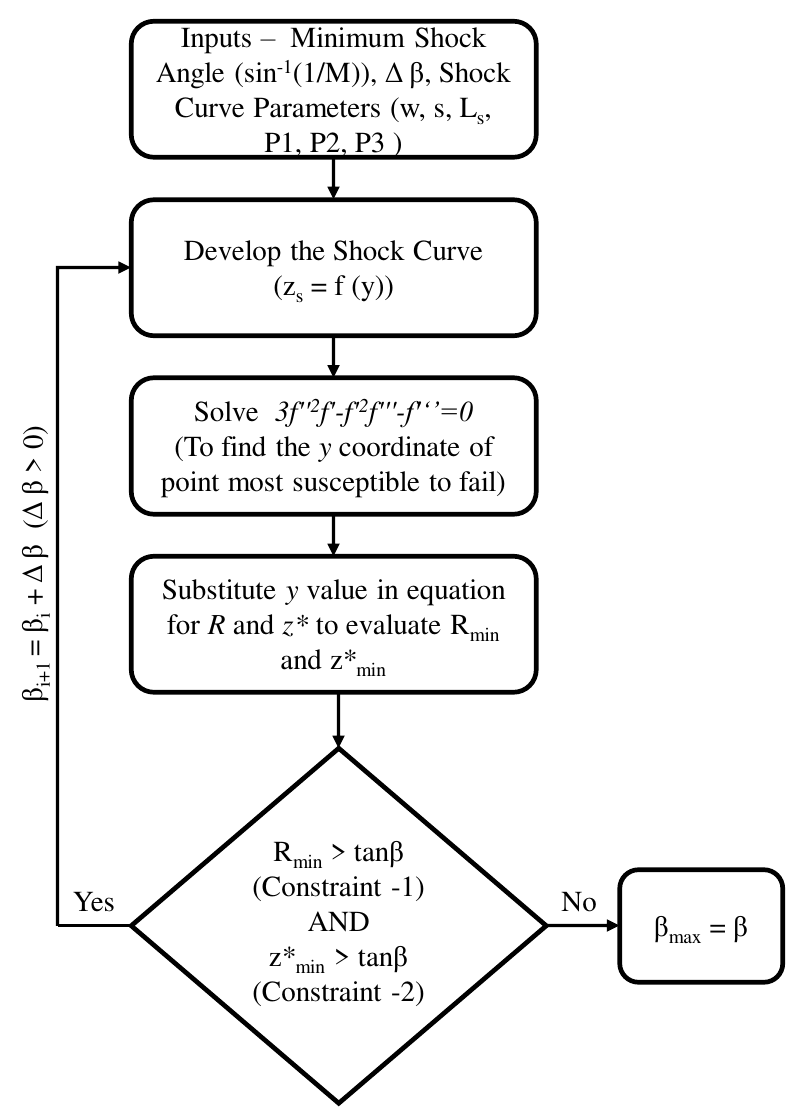}
    \caption{Flowchart showing the numerical framework for calculating maximum shock angle for Bézier Curve}
    \label{Numerical_Framework}
\end{figure}

As B\'{e}zier curves are parametric curves, it is difficult to derive an analytical expression like the one obtained in case of power law derived waveriders. Hence, a numerical framework was developed to find the maximum shock angle $(\beta_{max})$ possible. The constraints are solved by solving Equation [\ref{Final Equation 1},\ref{Final Equation 2}]. Then the critical point found is substituted in the expression of $R$[Equation \ref{Radius_Curvature_equation}] and $z^*$[Equation \ref{z*_Equation}] to find $R_{min}$ and ${z^*}_{min}$. A check procedure is carried out to see if $R_{min}$ and $z*_{min}$ is greater than $\tan\beta$. If the condition is satisfied $\beta$ is increased and the process continues until the condition fails. The  value of $\beta$ at which both Constraint 1 and Constraint 2 fail is maximum permissible shock angle$(\beta_{max})$ for the selected shock curve parameters . The framework is shown  as a flowchart in Figure[\ref{Numerical_Framework}].

The B\'{e}zier shock curve is defined by the control points $P_1$,$P_2$ and $P_3$. To analyse the effect of design parameters on the maximum shock angle, $P_3^*(P_3/w)$ is independently varied between 0.7 and 0.8 and $P_1$ and $P_2$ are equidistant points between $P_0$ and $P_3$. Hence, given a value of $L_s$ , $P_3$ and $P_4$ the shock curve is defined. The length ($L$) and semi span ($w$) of the waverider is fixed at 1 and 0.4 respectively to ensure aspect ratio $(w/L)$ of 0.4.

\begin{enumerate}
    \item \underline{Case-1}- Figure [\ref{Bezier_Graph_1}] shows the effect of control point 3 ($P_3$) for 3 different cases of tip height ($s$).The maximum permissible shock angle $(\beta_{max})$ decreases with an increase in the $y$ coordinate of $P_3$ for all the cases of tip height $s$. Moreover, a low value of $s$ will permit using higher shock angles for a specific value of $P_3$.
    \begin{figure}[hbt!]
        \centering
        \includegraphics[width=0.9\textwidth]{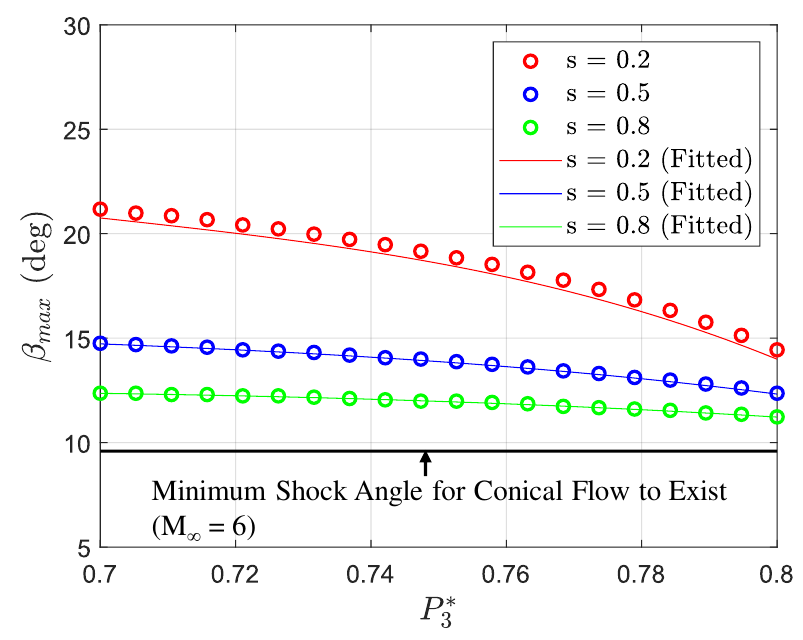}
        \caption{Maximum Permissible Shock Angle ($\beta_{max}$) vs $P_3^*$ for Different s}
        \label{Bezier_Graph_1}
    \end{figure}
    \item \underline{Case-2}- Figure [\ref{Bezier_Graph_2}] shows the effect of control point 3 ($P_3$) for different cases of $L_s$. The results obtained are similar to the ones obtained in case of power law.
    \begin{figure}[hbt!]
        \centering
        \includegraphics[width=0.9\textwidth]{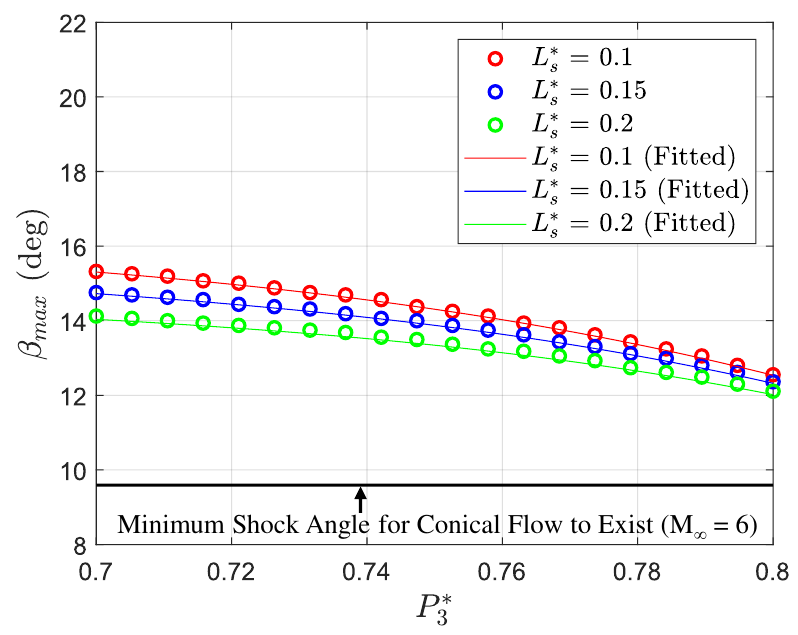}
        \caption{Maximum Permissible Shock Angle ($\beta_{max}$) vs $P_3^*$ for Different $L_s^*$}
        \label{Bezier_Graph_2}
    \end{figure}
\end{enumerate}
The numerical shock limit found by the above described framework is fitted using MATLAB's in built curve fitting toolbox. It is found that cubic polynomials provided the best fit to the numerical with root mean square error of less than 3\%. The corresponding coefficients of the fitted cubic polynomial for each case is given in Table [\ref{Fit_Coeff}].
The fitted cubic polynomials can be expressed as-
\begin{equation}
    y = C_0x^3 + C_1x^2 + C_2x +C_3
\end{equation}
Where, $y$ is the maximum shock angle ($\beta_{max}$) and $x$ is the location of $P_3^*$.
\begin{table}[hbt!]
\caption{\label{Fit_Coeff} Coefficients of the Fitted Cubic Polynomials}
\centering
\begin{tabular}{|c|c|c|c|c|c|c|}
\hline
Case & $s$ & $L_s^*$ & $C_0$ & $C_1$  & $C_2$ & $C_3$ \\
\hline
1 & 0.2 & 0.15 & -3169 & 6648 & -4684 & 1129\\
\hline
2 & 0.5 & 0.15 & -577.3 & 1158 & -785.4 & 195.1\\
\hline
3 & 0.8 & 0.15 & -148.5 & 260.6 & -151.3 & 41.51\\
\hline
4 & 0.5 & 0.1 & -469.3 & 901.5 & -586.8 & 145.3\\
\hline
5 & 0.5 & 0.2 & -526.8 & 1059 & -718.4 & 178.7\\
\hline
\end{tabular}
\end{table}
In the case of power law shock curve waveriders, it is evident that waveriders can be developed with higher shock angles. The rate of decrease in $\beta_{max}$ for power law waveriders is rather large, limiting the design space to some extent. Whereas,in case of B\'{e}zier shock curve, the drop in $\beta_{max}$ is substantially flatter, allowing for greater design flexibility. In case of power law shock curve, if $L_s$ and $s$ are fixed the shape of the shock curve changes according to $n$, whereas in case of Bezier shock curve the curve depends on $P_3^*$ hence, in some way $n$ and $P_3^*$ are analogous. For example, in Figures [\ref{Power_Graph_1},\ref{Bezier_Graph_1}] if $s$ is fixed at 0.5, it can be clearly seen that for power law, no waverider can be developed after $n=5.5$, whereas in case of Bezier shock curve, there is still scope of waverider generation for $P_3^* \geq 0.8$.
\section{Conclusions}
Conventionally low cone shock angles have been used in the design of osculating cone waveriders and the question whether there is a maximum limit of the cone shock angle for successful waverider generation remained unanswered. The principal aim of this study is to establish relationships that provides a maximum limit for the cone shock angle as a function of the design parameters of the shock curve. Such constraints will be useful to eliminate infeasible design parameters during automated multi disciplinary optimisation.

The shock curve has to satisfy two geometrical conditions for successful waverider geometry generation. The radius of curvature at any point should be greater than the distance between the shock curve and the base curve. Two osculating planes must not intersect between the shock curve and the base curve. Mathematical expressions defining the two constraints are derived for an arbitrary shock curve. Analysis is conducted for power law and Bezier based shock curves. Analytical expressions are obtained for the power law curve, while numerical procedure is developed for Bézier curve. Results for the Bézier curve are found to follow a cubic polynomial and the coefficients of the curve fit are evaluated.

Important conclusions are:
\begin{itemize}
    \item In all cases, the maximum cone shock angle$(\beta_{max})$ is significantly less than maximum angle for attached shock solutions.
    \item The maximum cone shock angle is highly sensitive to the design parameters of the shock curve.
    \item Power law curve shows higher maximum cone shock angles, but the value decreases rapidly with the power law exponent so much so that higher values of exponent fail to generate waverider geometry thereby limiting the design space.
    \item The trend remains the same for Bézier curve , however the rate of decrease is smaller enabling a larger design space.
\end{itemize}

\section*{Funding Sources}
The first author expresses gratitude to the Ministry of Education, Government of India, for the scholarship received.

\section*{Acknowledgements}
The authors would like to thank the members of Laboratory for Hypersonic and Shock Wave Research and Center of Excellence in Hypersonics, Indian Institute of Science for their support and meaningful discussions.

\appendix

\section{Detailed Constraint Development for a Generic Shock Curve}
\label{sec:sample:appendix}
Let the shock curve be a function $z=f(y)$.

According to \underline{Constraint 1}, $R_{min}>H_{max}$ i.e. the minimum radius of curvature of the shock curve must be greater than maximum height($H_{max}=Ltan\beta$).

Radius of Curvature($R$) is given by- 
\begin{equation}
    \label{Radius_Curvature_equation}
    R=\dfrac{(1+f'^2)^\frac{3}{2}}{|f''|}
\end{equation}
For maxima or minima of R,  $\dfrac{dR}{dy}$ is calculated  and equated  to 0. So differentiating the above equation with respect to $y$.
\begin{equation}
    3f''^2f'-f'''f'^2-f'''=0
    \label{Final Equation 1}
\end{equation}

According to \underline{Constraint 2}, no two consecutive osculating planes should intersect between the base curve and the shock curve.
The equation of the first osculating plane is-
\begin{equation}
    z-z_1=\dfrac{y-y_1}{-f'_1}
\end{equation}
The equation of the second osculating plane is- 
\begin{equation}
    z-z_2=\dfrac{y-y_2}{-f'_2}
\end{equation}
The $z$ coordinate intersection point($z^*$) is given by-
\begin{equation}
    z^*=\dfrac{(y_2-y_1)+z_2f'_2-z_1f'_1}{f'_2-f'_1}
\end{equation}
Let $y_1=y$ and $y_2=y_1+dy$,$y_2-y_1=dy$,
\begin{equation}
    \begin{split}
        &f'(y)=\dfrac{f(y+dy)-f(y)}{dy}\\
        &f(y+dy)=f'dy+f\\
    \end{split}
\end{equation}
Similarly, 
\begin{equation}
    f'(y+dy)=f''dy+f'
\end{equation}
The final equation of $z^*$ becomes,
\begin{equation}
    \label{z*_Equation}
    z^*=\dfrac{1+f'^2}{f''}+f\\
\end{equation}
We have to find the minimum value of $z^*$ and ensure that it is greater than $H_{max}$ for a feasible design. So we need to find $\dfrac{dz^*}{dy}$ and equate it to 0. 
\begin{equation}
    \dfrac{dz^*}{dy}=3f''^2f'-f'^2f'''-f'''=0
    \label{Final Equation 2}
\end{equation}
 \bibliographystyle{elsarticle-num-names} 
 \bibliography{references}





\end{document}